\documentclass{sig-alt-full}

\def\bigl{\left({\vrule height1.29em width0em depth1.29em}}
\def\bigr{{\vrule height1.29em width0em depth1.29em}\right)}

\def\ni{\noindent }
\def\eq #1{(\ref{#1})}       % Eq(N)
\def\l{\left}                   % big (
\def\r{\right}                  % big )
\newcommand{\be}[1]{\begin{equation}\label{#1}}
\def\ee{\end{equation}}

\newcommand{\ba}[1]{\begin{array}{#1}}
\def\ea{\end{array}}

\def\fr #1#2{\frac{#1}{#2}}

\def\se #1{sec.\,\ref{#1}}

\def\y1{\mbox{$y'$}}
\def\yt{\mbox{$y''$}}

\def\e{{\rm e}}

\def\hyper3{{\bf hyper3}}   

\def\2F1{\mbox{$_2${F}$_1$}}
\def\1F1{\mbox{$_1${F}$_1$}}
\def\0F1{\mbox{$_0${F}$_1$}}
\def\PFQ{\mbox{$_p${F}$_q\;$}}
\begin{document}

% --- Author Metadata here ---

\conferenceinfo{ISSAC'04,} {July 4--7, 2004, Santander, Spain.} 
\CopyrightYear{2004} 
\crdata{1-58113-827-X/04/0007} 

% --- End of Author Metadata ---

\title{Non-Liouvillian Solutions for Second Order Linear ODEs}

% \subtitle{[Extended Abstract]
% \titlenote{A full version of this paper is available as
% \textit{Author's Guide to Preparing ACM SIG Proceedings Using
% \LaTeX$2_\epsilon$\ and BibTeX} at
% \texttt{www.acm.org/eaddress.htm}}}

\numberofauthors{2}

\author{
\alignauthor L. Chan \\
       \affaddr{Department of Pure Mathematics}\\
       \affaddr{University of Waterloo}\\
       \affaddr{Waterloo, Ontario, Canada, N2L 3G1}\\
       \email{kclchan@pythagoras.math.uwaterloo.ca}
\alignauthor E.S. Cheb-Terrab\\
       \affaddr{CECM, Department of Mathematics}\\
       \affaddr{Simon Fraser University}\\
       \affaddr{Vancouver, British Columbia, Canada, V5A 1S6}\\
       \affaddr{Maplesoft, Waterloo Maple Inc.}\\
       \affaddr{Waterloo, Ontario, Canada, N2V 1K8}\\
       \email{ecterrab@cecm.sfu.ca}
}

\date{24 April 2004}
\maketitle

\begin{abstract}

There exist sound literature and algorithms for computing Liouvillian
solutions for the important problem of linear ODEs with rational
coefficients. Taking as sample the 363 second order equations of that type
found in Kamke's book, for instance, 51\% of them admit Liouvillian
solutions and so are solvable using Kovacic's algorithm. On the other hand,
special function solutions not admitting Liouvillian form appear frequently
in mathematical physics, but there are not so general algorithms for
computing them. In this paper we present an algorithm for computing special
function solutions which can be expressed using the \2F1, \1F1 or \0F1
hypergeometric functions. The algorithm is easy to implement in the
framework of a computer algebra system and systematically solves 91\% of the
363 Kamke's linear ODE examples mentioned.

\end{abstract}

%--------------------------------------------------------------------
% A category with the (minimum) three required fields

% \category{G.4}{Mathematical software}{Algorithm design and analysis.}
% A category including the fourth, optional field follows...
\category{I.1}{Symbolic and algebraic manipulation}{Algorithms.}
\terms{Algorithms, design, theory.}

\keywords{Linear ordinary differential equations, Non-Liouvillian solutions, hypergeometric solutions.}

%--------------------------------------------------------------------
\section*{Introduction}

Given a second order linear ODE
\be{LinearODE}
y'' + A(x)\,y' + B(x)\,y = 0
\ee

\ni where the quantity\footnote{This quantity is an invariant under
transformations of the dependent variable - see \eq{Inv}.} $A'/2+A^2/4-B$ is a
rational function of $x$, the problem under consideration is that of systematically
computing solutions for this ODE even when the solutions admit no Liouvillian
form\footnote{Functions that can be expressed in terms of exponentials,
integrals, and algebraic functions, are called Liouvillian functions. The
typical example is $\exp(\int R(x), dx)$ where $R(x)$ is rational or an
algebraic function representing the roots of a polynomial.}. 

The first thing to note is that non-Liouvillian solutions which are
representable symbolically not as unknown infinite sums can be
represented using special functions, e.g. Bessel, Hermite or Legendre
functions \cite{Seaborn}. In turn, these and most of the special functions
frequently appearing in mathematical physics happen to be particular cases
of the \PFQ hypergeometric function for $p$ equal to 0, 1 or 2 and $q$ equal
to 0 or 1 (see \cite{abramowitz}). For example, the Bessel functions can be
expressed in terms of \0F1, all cylindrical functions as well as the
Hermite, Laguerre, Whittaker and error family of functions can be expressed
in terms of \1F1, and all Chebyshev, Gegenbauer, Jacobi, Legendre and some
others can be expressed in terms of $\2F1$. 

% Overall, most of the mathematical functions
% frequently appearing in mathematical modeling can be expressed in terms of
% ese \PFQ functions.

One natural approach is then to directly attempt the computation of
hypergeometric function solutions of these \0F1, \1F1 and \2F1 types, since
in this way we cover at once solutions involving all the related special
functions. Such an approach was developed during the year 2001 (see
\cite{nserc}), it became the main algorithm of the Maple computer algebra
system for this type of problem since then and it is the subject of this
paper. The algorithm consists of an equivalence approach to the \PFQ
differential equations, is formulated in \se{hyper3}, 2 and 3, and computes
solutions of the form

\be{sol_hyper3}
y = P(x)\ \PFQ\l(..; ..; \fr{\alpha\, x^k + \beta}{\gamma\, x^k + \delta}\r)
\ee

\ni where $P(x)$ is an arbitrary function and
$\{\alpha,\beta,\gamma,\delta,k\}$ are constants.

It is important to note that the idea of seeking hypergeometric function
solutions for \eq{LinearODE} or using an equivalence approach for that is
not new. In '89 Kamran and Olver \cite{kamran} showed how to use an
equivalence approach to compute Bessel function solutions to eigenvalue
problems. Hypergeometric solutions were also discussed by Petkovsek and Salvy
\cite{petkovsek} in '93. Some of the more recent developments were
presented as computer algebra algorithms too. For instance, a classic
invariant theory approach was presented during 2000 by von B\"ulow in
\cite{kathi}; in 2001 Willis \cite{willis} presented a semi-heuristic
algorithm for computing special functions solutions. In 2002 Bronstein and
Lafaille \cite{bronstein} presented an approach for resolving an
equivalence under rational transformations, between two linear equations in normal form,
whenever one of them has an irregular singularity\footnote{That also leads to  \1F1
solutions of the form \eq{sol_hyper3}, including its particular \0F1 case, whenever the point of application of \1F1 is rational
in the independent variable.}.

There is natural intersection between what these algorithms can solve but
none can claim to extensively cover the portions of the problem covered by
all the others. If compared with the algorithm presented in this paper - we
called it hyper3 - these other algorithms, both those developed before and
after hyper3:

\begin{itemize}

\item Do not resolve in a systematic manner all of the \2F1, \1F1 and \0F1
equivalences;

\item Do not handle the problem of an invariant involving fractional or
abstract powers;

\item Do not explore automorphisms to avoid uncomputed integrals
in the solution. 

\end{itemize}

Also, hyper3 does not require solving systems of
algebraic equations nor computing Groebner basis nor running differential
elimination processes nor eliminating parameters by composing resultants (all of them expensive computational processes),
thus resulting in a fast and smooth algorithm with little computational
cost. These facts, combined with the range of problems it solves, for
instance taking Kamke's book \cite{kamke} as a testing arena, are at the
base of the role hyper3 has today in the Maple differential equation
libraries.

%---------------------------------
\section{Computing \2F1, \1F1 and \0F1 hypergeometric solutions}
\label{hyper3}

To compute \PFQ solutions to \eq{LinearODE}, the idea is to formulate an
equivalence approach to the \PFQ underlying hypergeometric
differential equations; that is, to determine whether a given linear ODE can
be obtained from one of the \2F1, \1F1 or \0F1 ODEs, respectively given by

\be{seeds}
\begin{array}{rcl}
\left( {x}^{2}-x \right) \yt
+ \left(  \left(  a + b +1 \right) x- c 
\right) \y1 +  b \, a \,y  & = & 0,\ \ % \mbox{(\2F1)}
\\
x\yt + \left(  c  - x \right) \y1  -  a \,y & =  & 0,\ \ % \mbox{(\1F1)} 
\\
x\yt +  c  \y1 -y  & =  & 0, %,\ \ \mbox{(\0F1)}  
\end{array}
\ee

\ni where $\{a,b,c\}$ are arbitrary constants, by means of a transformation
of a certain type. If so, the solution to the given linear ODE is obtained
by applying the same transformation to the solution of the corresponding
\PFQ ODE above.

This approach of course also requires determining the values of the
hypergeometric parameters $\{a, b, c\}$ for which the equivalence exists, and it
is clear that its chances of success depend crucially on how general is the
class of transformations being considered. For instance, one can verify that
for linear transformations\footnote{The problem of equivalence under
transformations $\{x \rightarrow F(x),\ \ y \rightarrow P(x)\, y +
Q(x)\}$ for linear ODEs can always be mapped into one with $Q(x) = 0$, see
\cite{ince}.} 

\begin{equation}
x \rightarrow F(x),\ \ y \rightarrow P(x)\, y
\ee

\ni with arbitrary $F(x), P(x)$, the problem is too general in that to solve
it requires solving first the given ODE, so that the approach is of no
practical use \cite{kathi}. 

The transformations considered in this work are 

\be{tr}
x \rightarrow \fr{\alpha\, x^k+\beta}{\gamma\, x^k+\delta},\ \ y \rightarrow P(x)\,y
\ee

\ni with $P(x)$ arbitrary and $\{\alpha, \beta, \gamma, \delta, k\}$
constant with respect to $x$. These transformations, which do not conform a
class in the strict sense\footnote{By class of transformations we mean a set of transformations closed under composition.}, can be obtained by sequentially composing three
different transformations each of which does constitute a class. The sequence
starts with linear fractional - also called M\"obius - transformations

\be{mobius}
M := x \rightarrow \fr{\alpha\, x+\beta}{\gamma\, x+\delta},
\ee

\ni is followed by power transformations

\be{power}
x \rightarrow x^k,
\ee

\ni and ends with linear homogeneous transformations of the dependent variable

\be{linear_in_y}
y \rightarrow P\,y
\ee

\ni So, we are talking of an algorithm that systematically computes, when they
exist, solutions of the form

\be{sol_pattern}
y = P(x)\ \PFQ\l(..; ..; \fr{\alpha\, x^k + \beta}{\gamma\, x^k + \delta}\r)
\ee

\ni where \PFQ is any of \2F1, \1F1 or \0F1.

\subsection{Transformations $y \rightarrow P(x)\, y$ of the dependent variable}
\label{y_equals_P_times_u}

The first thing to note is that transformations of the form \eq{linear_in_y}
can easily be factored out of the problem: if two equations of the form
\eq{LinearODE}, with coefficients $\{A(x),B(x)\}$ and
$\{C(x),D(x)\}$ respectively, can be obtained from each other by means of
\eq{linear_in_y}, the transformation relating them is computable 
from these coefficients. For that purpose, we rewrite both equations in
normal form, for instance for \eq{LinearODE} use 

\be{ToNormalForm}
y=u\,{e^{-\int \!A/2\,{dx}}}
\ee

\ni to obtain

\be{NormalForm}
{\it u''}= \left( \fr{A'}{2} + \fr{{A}^{2}}{4}\ - B \right) u
\ee

\ni and the transformation relating the two hypothetical ODEs exists when the
two normalized equations are equal; the transformation relating them being
$y=u\,{e^{\int \!(C-A)/2\,{dx}}}$. In what follows we will refer to 

\be{Inv}
I(x) = \fr{A'}{2} + \fr{{A}^{2}}{4}\ - B,
\ee

\ni the coefficient of $u$ in \eq{NormalForm}, as {\em the invariant}
\cite{olver}, regardless of the fact that this object is only an absolute
invariant under \eq{linear_in_y} and not under \eq{mobius} or \eq{power}.

\subsection{Transformations $x \rightarrow F(x)$ of the independent variable}
\label{classification}

By changing $x \rightarrow F(x)$ in \eq{LinearODE}, the invariant $I_1$
of the changed ODE can be expressed in terms of the invariant $I_0$ of
\eq{LinearODE} by 

\be{Inv_1}
I_1(x) =F'^{2}I_0(F(x)) + S(F') 
\ee

\ni where $S(x)$ is the Schwarzian \cite{weisstein}
\be{Schwarzian}
S(F') = \fr{3 F''^2}{4 F'^2} - \fr{F'''}{2 F'};
\ee

\ni The form of $S(F')$ is particularly simple when $F(x)$ is a power
transformation (see \eq{S_xk}) and also when $F(x)$ is a M\"obius transformation
\eq{mobius}, in which case $S(F') = 0$. These are key facts permitting a
simple formulation and resolution of the equivalence.

%---------------------------------
\section{M\"obius transformations and a classification of singularities}

The first ODE in \eq{seeds} has 3 regular singularities, at $0$, $1$ and
$\infty$. The second ODE in \eq{seeds}, also known as the confluent
hypergeometric equation, has a regular singularity at $0$ and an irregular
one at $\infty$. The third ODE in \eq{seeds} also has one regular and one
irregular singularity at $0$ and $\infty$, but we considered the case
separately in order to obtain solutions directly expressed in terms of
simpler (Bessel) functions. As we shall see, the structure of the
singularities of these equations is a key for resolving related
equivalences and M\"obius transformations preserve
that structure. These transformations only move the location of the poles.
For example, the \0F1 hypergeometric equation

\be{0F1}
x\,\yt +  c\,  \y1 - y  =  0
\ee

\ni has one regular singularity at the origin and one irregular at infinity.
The transformed ODE, obtained from \eq{0F1} by means of \eq{mobius}

% \begin{equation}
% \yt + {\frac { \left( 2\,\alpha\,\gamma\,x+2\,\gamma\,\beta-
% \gamma\,c\beta+c\alpha\,\delta \right) }{ \left( \alpha\,x+
% \beta \right)  \left( \gamma\,x+\delta \right) }\,\y1}
% - {\frac { \left( 
% \alpha\,\delta-\gamma\,\beta \right) ^{2}}{ \left( \gamma\,x+\delta
%  \right) ^{3} \left( \alpha\,x+\beta \right) }\,y} = 0
% \ee

\begin{eqnarray}
\nonumber
\yt 
+ 
\lefteqn{
{\frac {\left( \alpha\, \left( \delta\,c+2\,\gamma\,x \right) +\gamma\, \left( 2-c \right) \beta \right) }{ \left( \alpha\,x+\beta \right)  \left( \gamma\,x+\delta \right) }}\, \y1
}
&  & 
\\
& - & 
\!\!\!{\frac { \left( \alpha\,\delta-\gamma\,\beta \right) ^{2}}{ \left(\gamma\,x+\delta \right) ^{3} \left( \alpha\,x+\beta \right) }}\, y  =  0 
\end{eqnarray}

\ni also has one regular and one irregular singularity, respectively located
at $-{{\beta}/{\alpha}}$ and $-{{\delta}/{\gamma}}$. In the case of the \2F1
equation (see \eq{seeds}), under \eq{mobius} the three regular singularities
move from $\{0,1,\infty\}$ to $\{-\delta/\gamma, -\beta/\alpha,
(\delta-\beta)/(\alpha-\gamma)\}$. So, from the structure of the
singularities of an ODE, not only one can tell with respect to which of the
three differential equations \eq{seeds} could the equivalence under
\eq{mobius} be resolved, but also one can extract information regarding the
values of the parameters $\{\alpha, \beta, \gamma, \delta\}$ entering the
transformation.

Reversing the line of reasoning, through Mobius transformations one can
formulate a classification of singularities of the linear ODEs ``equivalent"
to the \PFQ equations \eq{seeds}, based on how the invariant of each of
these equations is transformed. Concretely, after transforming the \2F1
equation, the invariant of the resulting equation has the form

\be{Inv2F1}
I_{{{\2F1}}} = 
{\frac {\omega_{{2}}{x}^{2}+2\,\omega_{{1}}x+\omega_{{0}}}{ \left( 
\sigma_{{1}}x+\sigma_{{2}} \right) ^{2} \left( \sigma_{{3}}x+\sigma_{{
4}} \right) ^{2} \left( \sigma_{{5}}x+\sigma_{{6}} \right) ^{2}}}
\ee

\ni where all $\{\omega_i, \sigma_j\}$ can be expressed in terms of
$\{a,b,c\}$ and $\{\alpha,\beta,\delta,\gamma\}$ respectively entering the
\2F1 equation \eq{seeds} and the transformation \eq{mobius}. The invariant
of the transformed \1F1 equation has the form

\be{Inv1F1}
I_{{{\1F1}}} = 
{\frac {\omega_{{2}}{x}^{2}+2\,\omega_{{1}}x+\omega_{{0}}}{\left( \sigma_{{3}}x+\sigma_{{
4}} \right) ^{2} \left( \sigma_{{5}}x+\sigma_{{6}} \right) ^{4}}}
\ee

\ni and that of the transformed \0F1 equation has the form

\be{Inv0F1}
I_{{{\0F1}}} = 
{\frac {\omega_{{1}}x+\omega_{{0}}}{\left( \sigma_{{3}}x+\sigma_{{
4}} \right) ^{2} \left( \sigma_{{5}}x+\sigma_{{6}} \right) ^{3}}}
\ee

\ni These transformed invariants are all of the form

\be{GenericInvariant}
I_{\PFQ}={\frac {\prod_{i=1}^m \left( a_i x + b_i\right) }{\prod _{i=1}^{n} \left( c_{{i}}x+d_{{i}} \right) ^{q_{{i}}}}}
\ee

\ni Cancellations between factors in the numerator and denominators of
\eq{GenericInvariant} may also happen and, independent of that, some
coefficients $\{a_i,c_i\}$ can be zero\footnote{Provided that, in
\eq{mobius}, $\alpha \delta - \gamma \beta \neq 0$ and also that in
\eq{LinearODE} the invariant remains finite, i.e. its denominator is not
zero.}. So the degrees with respect to $x$ of the numerators and
denominators of \eq{Inv2F1}, \eq{Inv1F1} and \eq{Inv0F1} can be lower than
the maximum implicit by these equations; in this way the problem splits into
cases.

Taking these possible cancellations into account, from the structure of the
invariants \eq{Inv2F1}, \eq{Inv1F1} and \eq{Inv0F1}, the different cases
for each of the \2F1, \1F1, \0F1 classes were determined. With this
classification in hands, from the knowledge of the degrees with respect to
$x$ of the numerator and denominator of the invariant \eq{GenericInvariant}
of a given ODE, one can tell whether or not it can be obtained from the
\2F1, \1F1 or \0F1 equations \eq{seeds} using \eq{mobius}. These
observations can be summarized in a classification table as follows, using
the symbol

$$[\leq p,\, [q_1 *,q_2 *,...,q_n *]]$$

\ni where $p$ is the degree in $x$ of the numerator of \eq{GenericInvariant}
and $q_i$ are the powers of the factors entering the denominator of it. The
symbol $\leq$, when present, refers to the value of $p$ (can be less or
equal to). The symbol $*$, when present, means there can be factors
canceling between numerator and denominator, so that the actual value of the
related $q_i$ can be lower (provided $p$ is also lower by the same amount).
For example, 

\begin{equation}
[\leq 2*, [2*, 2*]] 
\ee

\ni represents the following possible seven different ``lists of values" (herein
referred as cases) for the degrees of the numerator and denominator of the
invariant

\begin{equation}
\begin{array}{rcl}
[2*, [2*, 2*]] & = &  [2, [2, 2]], [1, [1, 2]], [0, [1, 1]], [0, [0, 2]]
\\*[0in]
[1*, [2*, 2*]] & = &  [1, [2, 2]], [0, [1, 2]] 
\\*[0in]
[0, [2, 2]]
\end{array}
\end{equation}

\ni With this notation, the classification of all the possible cases
equivalent to the \2F1, \1F1 and \0F1 equations under M\"obius
transformations is as shown in Table 1.

\begin{table*}
{\begin{center} {% \footnotesize
\begin{tabular}{|c|l|c|}
\hline

    Class  &  Cases & Number of cases \\
    
\hline 

    \2F1 & $[<=2*, [2*, 2*, 2*]], [<=2*, [2*, 2*]]$ &  14 \\
    
\hline 

    \1F1 & $ [2*, [2*, 4]], [<=2, [6]], [<=2, [4]], [2*, [2*]], [2, [0]]$ & 13 \\
    
\hline

    \0F1 & $[1*, [2*, 3]], [<=1, [5]], [<=1, [3]], [1*, [2*]], [1, [0]]$ & 9 \\
    
\hline
\multicolumn{3}{c}
       {Table 1. \it Classification of linear ODEs equivalent to \PFQ ODEs under M\"obius}\\
\end{tabular}}
\end{center}}
\end{table*}

%---------------------------------
\vfill\eject 
\section{Transformations $x \rightarrow x^k$ of the independent variable}

Using the results of the previous sections it is possible to resolve the
equivalence of a given linear ODE \eq{LinearODE} and the hypergeometric
equations \eq{seeds} under compositions of transformations \eq{linear_in_y}
of the dependent variable $y(x)$ and M\"obius transformations \eq{mobius} of
the independent variable $x$. In this section a worth additional level of
generalization is obtained by composing those two transformations with
transformations $x \rightarrow x^k$ of the independent variable.

The first thing to note regarding power transformations is that, unlike
M\"obius transformations, they {\em do not} preserve the structure of
singularities. The change in the invariant due to $x \rightarrow x^k$,
however, has a simple and tractable structure. The Schwarzian
\eq{Schwarzian} is given by:

\be{S_xk}
S(F') = {\frac {{k}^{2}-1}{4\, {x}^{2}}}
\ee

\ni So, the changed invariant $I_1$ shown in \eq{Inv_1} can be expressed in
terms of $I_0$ by

\begin{equation}
{x}^{2} I_{{1}}( x ) + \fr{1}{4} = 
    \left(  \left( {x}^{k} \right)^{2} I_{{0}}( {x}^{k}) + \fr{1}{4} \right) {k}^{2}
\ee

\ni This naturally suggests the introduction of a ``shifted" invariant
$J(x)$

\be{J}
J_{{i}}( x ) ={x}^{2}I_{{i}}( x ) +\fr{1}{4}
\ee

\ni for which the transformation rule under $x \rightarrow x^k$ has the simple
form

\be{J1J0}
J_{{1}}( x ) ={k}^{2}J_{{0}} ( {x}^{k} ) 
\ee

The equivalence of two linear ODEs A and B under $x \rightarrow x^k$ can
then be formulated as follows: Given $J_{1A}(x)$ and $J_{1B}(x)$, compute
$k_A$ and $k_B$ entering \eq{J1J0} such that the degrees with respect to
$x$ of $J_{0A}(x)$ and $J_{0B}(x)$ are minimized. This approach is
systematic: equations A and B are related through power transformations only
when $J_{0A} = J_{0B}$ and, if so, the mapping relating A and B is just $x
\rightarrow x^{k_A-k_B}$.

The computation of $k$ minimizing the degrees of $J_0$ in \eq{J1J0} is
formulated as follows. Given the set 

\be{k}
A := \fr{p_i}{q_i},\ \ \ \ \ \mbox{$i = 1$ to $m$}
\ee

\ni of (possibly rational) numbers entering as exponents in the powers of
the independent variable found in $J_1$, compute the smallest rational
number $\tilde{k}$ such that multiplying by it each element of $A$, all of
them become integers. Then the value of $k$ minimizing the degrees of $J_0$
is $k = 1/\tilde{k}$.

%---------------------------------
\vfill\eject 
\section{Summary of hyper3 - examples}

An itemized description of the algorithm, discussed in the previous
subsections to resolve the equivalence proposed in the introduction, is as
follows.

\begin{enumerate}

\item Rewrite the given equation \eq{LinearODE} we want to solve in normal
form

\begin{equation}
y'' = I(x)\, y
\ee

\ni where $I(x)$ is the invariant \eq{Inv}.

\item Compute $J_1(x)$, the shifted invariant \eq{J}, and use
transformations $x\rightarrow x^k$ to reduce to the integer minimal values
the exponents of powers entering $J_0(x)$; i.e., compute $k$ and with it
compute $J_0(x)$ in \eq{J1J0}.

\item From \eq{J}, compute $I_0(x)$ and classify its structure of singularities according to
Table 1, to tell whether an equivalence under M\"obius transformations is
possible and to which of the \2F1, \1F1 or \0F1 equations \eq{seeds}.

\item When the equivalence is possible, from the singularities of $I_0(x)$
and by comparing it with the invariant \eq{GenericInvariant} of the transformed \PFQ
equation\footnote{At this point, $J_0(x)$ and the shifted invariant of the
\PFQ equation have the same degrees.}, compute the parameters $\{a, b, c\}$
entering the \PFQ equation \eq{seeds} such that the equivalence exists as
well as the parameters $\{\alpha, \beta, \gamma, \delta\}$ entering the
M\"obius transformation \eq{mobius}.

\item Compose the three transformations to obtain one of the form

$$
x \rightarrow \fr{\alpha x^k+\beta}{\gamma x^k+\delta},\ \ y \rightarrow P(x)\,y
$$

mapping the \PFQ equation involved into the ODE being solved.

\item Apply this transformation to the known solution of the \PFQ equation
resulting in the desired ODE solution.

\end{enumerate}

\medskip
\ni{\bf An example of the \2F1 class}
\medskip

Consider the second order linear ODE

% yY( _y2 = (2*(nu-mu)*x^2-3*x^4 - 2*(mu+nu)-1)/(x^5-x)*_y1 + nu*(nu+2*(mu+1))/(x^6-x^2)*y )

\begin{eqnarray}
\label{e1}
\nonumber
y'' \! & =  & \!
    {\frac { 2 \left( \nu - \mu \r){x}^{2}-3\,{x}^{4} -2\,(\mu+\nu)-1}{{x}^{5}-x}\, y'}
\\
& & 
+\, {\frac {\nu\, \left( \nu + 2\,(\mu+1) \r)}{{x}^{6}-{x}^{2}}\,y}
\end{eqnarray}

\ni This equation has regular singularities at $\{0, 1, -1, i, -i\}$.
Following the steps outlined in the Summary, we rewrite the equation in
normal form and then compute the value of $k$ leading to an equation
with minimal degrees for the powers entering $J_0(x)$ in \eq{J1J0}. The value
found is $k = 2$. So, using\footnote{This transformation is the
composition of $t\equiv x^k=x^{2}$ with a transformation of the form
\eq{ToNormalForm} so that \eq{e11} is normalized.}

% {t = x^2, u(t) = x^(1/2)*y(x)*exp(-1/2*Int((-3*x^4+(2*nu-2*mu)*x^2-2*nu-1-2*mu)/(x^5-x),x))}

\be{e11_to_e1}
t={x}^{2},\ \ 
u=\sqrt {x}\,{\e^{\l( %\displaystyle 
\int \!{\frac { 2\left( \nu - \mu \r) {x}^{2}-3\,{x}^{4} - 2\,(\mu+\nu)-1} {2\,(x-{x}^{5})} }{dx}\r)}}
\, y,
\ee

\ni the given equation \eq{e1} can be obtained from 

% diff(diff(u(t),t),t) = 1/4*((-2*mu+mu^2+2*nu^2-4)*t^2+(-2*nu^2+2*mu^2)*t+2*mu+mu^2)/(t^4-2*t^2+1)/t^2*u(t)

\be{e11}
u''={\frac { \left( {\mu}^{2} + 2\,({\nu}^{2} -\mu -2) \r) {t}^{2}+ 2\left( {\mu}^{2}-{\nu}^{2} \r) t
+ \mu\,({\mu}+2)} {4\, {t}^{2} (t-1)^2\,(t+1)^2 }}
\, u,
\ee

\ni which is in normal form and has an invariant with ``minimal
degrees" with respect to power transformations \eq{power}. 

In step 3, analyzing the invariant of \eq{e11} (coefficient of $u$ in its
right-hand-side), the equation has now three regular singular points, at
$\{0, 1, -1\}$. Using the notation of \se{classification}, the degrees with
respect to $t$ of the numerator and of each of the linear factors entering
the denominator are $[2,[2,2,2]]$. The equation matches the
classification Table 1 presented in \se{classification} and is identified as
equivalent to the \2F1 equation under M\"obius transformations \eq{mobius}.

So we proceed with step 4, equating the invariant of \eq{e11} with the
invariant \eq{Inv2F1} written in terms of
$\{a,b,c,\alpha,\beta,\gamma,\delta\}$, from where
we compute the values of the hypergeometric parameters $\{a,b,c\}$ entering
the \2F1 equation \eq{seeds}, such that the equivalence under M\"obius
exists, as well as the M\"obius transformation itself, obtaining

% [_a = 1/2*nu, _b = -mu+1/2*nu, _c = -mu]
% {y = u(t)*exp(Int(-1/2*((-mu+2*nu+2)*t-mu)/t/(t^2-1),t))/((t-1)^(1/2-1/2*nu+1/2*mu))/((t+1)^(1/2*nu+1/2))*t^(1/2*mu), x = 2*t/(t-1)}

$$
\{a = \fr{\nu}{2},\ b = \fr{\nu}{2}-\mu,\ c = -\mu\}\ \ \ \ M := x = \fr{2\,t}{t-1}
$$

\ni The transformation mapping the \2F1 equation \eq{seeds} at these values
of the parameters $\{a,b,c\}$ into \eq{e11} is then obtained composing the
M\"obius transformation above with one of the form \eq{linear_in_y},
computed as explained in \se{y_equals_P_times_u}, resulting in

\be{tr_M_y}
x = \fr{2\,t}{t-1},\ \ \
y = 
\fr
    { {t}^{\mu/2}\,  \l( t-1 \r)^{(\nu-\mu-1)/2}} 
    {  \l( t+1 \r) ^{(\nu+1)/2} } \, u(t)
\ee

\ni At this point, we have the transformation \eq{tr_M_y}
mapping \eq{seeds} into \eq{e11}, and the transformation
\eq{e11_to_e1}, mapping \eq{e11} into the equation \eq{e1} we want to solve.
Composing these transformations, in step six we obtain the solution of
\eq{e1}

% y(x) = x^nu/(x^2-1)^(nu/2)*hypergeom([nu/2,nu/2-mu],[-mu], 2*x^2/(x^2-1))*_C1 + x^(nu+2*mu+2)/(x^2-1)^(1+mu+nu/2)*hypergeom([nu/2+1,1+mu+nu/2],[2+mu],2*x^2/(x^2-1))*_C2;

\begin{eqnarray}
\label{sol_e1}
\nonumber
y \!\! & = & \!\!
\fr{{x}^{\nu}}{ \l( {x}^{2}-1 \r) ^{\fr{\nu}{2}\,}}\,{\2F1\l(\fr{\nu}{2},\, \fr{\nu}{2}\,-\mu;\,-\mu;\,{\frac {2\,{x}^{2}}{{x}^{2}-1}}\r)}\, 
C_1 
\\*[0.1in]
& + &
\fr{{x}^{\nu+2\,\mu+2}} {\l( {x}^{2}-1 \r)^{1+\mu+\nu/2}}
\\
\nonumber
& &
{\2F1\l(\fr{\nu}{2}+1,\, 1+\mu+\fr{\nu}{2};\, 2+\mu; \,{\frac
{2\,{x}^{2}}{{x}^{2}-1}}\r)}\,
C_2 
\end{eqnarray}

\ni where $C_1$ and $C_2$ are arbitrary constants. 

As mentioned in the introduction, an implementation of the algorithm being
presented is at the core of the current Maple ability to solve this type of problem.
The time consumed by this Maple implementation to compute the solution
\eq{sol_e1} performing all the steps mentioned is 0.4 seconds in a Pentium
IV, 2 GigaHertz computer. The Maple command line to
compute this solution directly using hyper3 is: {\verb-> dsolve(ode,[hyper3]);-}.

\medskip
\ni{\bf An example of the \1F1 class}
\medskip

As an example which also requires an extension of the algorithm to handle
symbolic powers in the invariant \eq{Inv}, consider Kamke's second order
linear equation 2.15:

\be{k215}
\yt+ \l( \mu\,{x}^{2\,\sigma} + \nu\, {x}^{\sigma-1} \r) y=0
\ee

\ni where $\mu,\, \nu$ and $\sigma$ are constants with respect to $x$. This
equation is already in normal form and the shifted invariant \eq{J} for it
is

\be{Jk215}
J_1(x) = 1/4-{x}^{2} \l( \mu\,{x}^{2\,\sigma} + \nu\,{x}^{\sigma-1} \r)
\ee

To compute the values of $k$ entering \eq{J1J0} and leading to $J_0(x)$ with
minimized integer powers, in \eq{k}, instead of restricting $\tilde{k}$ to
be a rational number, we allow it to depend on symbolic variables. So we
compute $\tilde{k}$ such that the set of exponents entering \eq{Jk215}, $A:=
\{2\sigma+2,\, \sigma+1\}$, becomes a set of integers after multiplying each
element of it by $\tilde{k}$, resulting in\footnote{To perform this
computation, it suffices to sequentially take the {\bf gcd} between each of
the elements of A.} $\tilde{k} = 1/(\sigma+1)$. In summary, using $
\left\{t={x}^{\,\sigma+1},\,u(t)={x}^{\,\sigma/2}\,y(x) \right\}$, Kamke's
equation \eq{k215} can be obtained from the following equation, which is
already in normal form and has an invariant with
minimized integer degrees, free of symbolic powers

\be{transformed_k215}
{\it u''}=-{\frac { \l( 4\,\mu\,{t}^{2}+4\,\nu\, t +
{\sigma}^{2}+2\,\sigma\r)}{ 4\, \l( \sigma+1 \r) ^{2}{t}^{2}}}\,
u
\ee

Proceeding with step 3, the invariant is the coefficient of
$u$ in the above and the degrees with respect to $t$ of its numerator and
factors in its denominator match the Table 1 of \se{classification},
identifying \eq{transformed_k215} as equivalent to the \1F1 equation under
M\"obius transformations \eq{mobius}. 

As in the previous example, in step 4, comparing the invariant of
\eq{transformed_k215} with the invariant \eq{Inv1F1} of the transformed
\1F1 equation, we compute the values of the parameters entering
the \1F1 equation \eq{seeds} such that the equivalence exists, as well as
the parameters entering the M\"obius transformation. Composing all the
transformations, we arrive at the solution for Kamke's example 2.15

\begin{eqnarray}
\nonumber
y \lefteqn{\ =} & &
{\e^{\l(-\displaystyle{\frac {i\,\sqrt {\mu}\,{x}^{\sigma + 1}}{\sigma + 1}}\r)}}
\bigl \right.
\\
& & 
\1F1\l( 
    {\frac {\sqrt {\mu}\,\sigma+i\,\nu}{2\,\sqrt {\mu} \l( \sigma + 1\r) }}; 
    {\frac {\sigma}{\sigma + 1}}; 
    {\frac {2\,i\,\sqrt {\mu}}{\sigma + 1}}\,{x}^{\sigma + 1} \r)\,
C_1
\\*[0.1in]
\nonumber
& &
+ \left.
{\1F1 \l( 
    {\frac {\sqrt{\mu}\,(\sigma+2)+i\,\nu}{2\,\sqrt {\mu} \l( \sigma + 1 \r) }};
    {\frac {\sigma + 2}{\sigma + 1}};
    {\frac {2\,i\,\sqrt {\mu}}{\sigma + 1}}\,{x}^{\sigma + 1} \r) x\, C_2}
    \bigr
\end{eqnarray}

\ni where $C_1$ and $C_2$ are arbitrary constants. The time consumed by the
implementation in Maple to perform these steps and return the solution above
is again 0.4 seconds, as in the previous example. This also illustrates
that, for typical problems, the additional handling of symbolic powers does not imply on any important performance cost.

%---------------------------------
\section{On the computation of the second independent solution}

The algorithm presented is based on computing a transformation mapping a
\PFQ equation into a given linear ODE, then applying that transformation to
the solution of the \PFQ equation to obtain the solution for the given
problem. This process has a subtlety: depending on the values of the
hypergeometric parameters, we may have only one independent solution
available for the \PFQ equation. In these cases, the second independent
solution can be obtained through integration: if $y=S(x)$ is a
solution of \eq{LinearODE}, then

\be{sol2}
y=\int \!{\frac {{\e^{\l( \int \!{\it A(x)} {dx}\r)}}}{{S(x)}^{2}}}\,{dx}\,S(x)
\ee

\ni is a second independent solution directly computable from $S(x)$ and $A(x)$. 

This approach, however, frequently introduces uncompu\-ta\-ble integrals, thus
complicating further manipulations and undermining the usefulness of the
result. As an example of this situation, for the \2F1 equation,

\be{ode_2F1}
\l( {x}^{2}-x \r) \yt
+ \l(  \l(  a + b +1 \r) x- c 
\r) \y1 +  b \, a \,y   =  0,
\ee

\ni the two independent solutions are:

\begin{eqnarray}
\label{sol_2F1}
y & = & \2F1 \l( a,b;\,c;\,x \r)\, C_1
\\
\nonumber
& & 
+\ {x}^{1-c}\, \2F1 \l( b-c+1,a-c+1;\,2-c;\,x \r)\, C_2 
\end{eqnarray}

\ni but for $c=1$ these two solutions are equal. Using the integration
recipe \eq{sol2}, a second independent solution is

\be{solint}
y = \int \!\fr{\e^{\l( \displaystyle \int \!{\frac { \l( a+b+1 \r) x-1}{{x}^{2}-x}}\,{dx} \r)}} 
    { \2F1 \l( a,b;\,1;\,x \r) ^{2}}\,{dx}\ \2F1 \l( a,b;\,1;\,x \r) 
\ee

\ni Although the inner integral, with rational integrand, is easy to
compute, the outer integral, with $\2F1 \l( a,b;\,1;\,x \r)
^{2}$ in its denominator, is uncomputable in current computer algebra
systems.

The approach used in hyper3 to minimize the occurrence of uncomputable
integrals consists of exploring the group of automorphisms of the \2F1
equation in order to make $c$ not an integer when that is possible.
Recalling, the group elements and their action are

{\begin{center} {\footnotesize
\begin{tabular}{|l|c|}
\hline

    Group element  &  Action on the plane  \\
    
\hline 

    $g_1: x \rightarrow x$ & $(0 \rightarrow 0, 1 \rightarrow 1, \infty \rightarrow \infty)$ \\
    
\hline 

    $g_2: x \rightarrow 1-x$ & $(0 \rightarrow 1, 1 \rightarrow 0, \infty \rightarrow \infty)$ \\

\hline 

    $g_3: x \rightarrow 1/x$ & $(0 \rightarrow \infty, 1 \rightarrow 1, \infty \rightarrow 0)$ \\

\hline 

    $g_4: x \rightarrow 1/(1-x)$ & $(0 \rightarrow 1, 1 \rightarrow \infty, \infty \rightarrow 0)$ \\

\hline 

    $g_5: x \rightarrow (x-1)/x$ & $(0 \rightarrow \infty, 1 \rightarrow 0, \infty \rightarrow 1)$ \\

\hline 

    $g_6: x \rightarrow x/(x-1)$ & $(0 \rightarrow 0, 1 \rightarrow \infty, \infty \rightarrow 1)$ \\
\hline
\multicolumn{2}{c}
       {Table 2. \it Group of automorphisms of the \2F1 equation}\\
\end{tabular}}
\end{center}}

\ni These transformations, known to act as permutations on the set
$\{0,1,\infty\}$, also act as permutations on a set $\{\lambda, \mu,
\kappa\}$ related to the hypergeometric parameters $\{a, b, c\}$ by 

\begin{equation}
\lambda = 1-c,\ \ \mu = a + b - c,\ \  \kappa = a - b
\ee

\ni These three parameters are the exponent differences of the normal form
of the \2F1 equation \eq{seeds}, at $\{0,1,\infty\}$ respectively. The
action of each $g_i$ on these parameters is obtained from Table 2 by
respectively changing $\{0, 1, \infty\}$ by $\{\lambda, \mu, \kappa\}$.
Hence, the solution \eq{sol_2F1} can be written in different manners, by
changing the application point of the \2F1 function using the $g_i$,
permuting accordingly the parameters $\{\lambda, \mu, \kappa\}$ entering the
\2F1 function and multiplying the result by the proper non-constant
factor\footnote{These multiplicative factors are different for each $g_1$;
we omit them here for brevity.}.

For example, when $c$ is an integer but $a+b$ is not an integer, applying
$g_2$ and permuting the parameters $\mu \leftrightarrow \lambda$, the
power $x^{1-c}$ entering \eq{sol_2F1} becomes a power with non-integer
exponent. Using this mechanism, for \eq{ode_2F1} at $c=1$, instead of the
solution with integrals \eq{solint} we obtain two independent solutions free
of uncomputed integrals:

\begin{eqnarray}
y \!\! & = & \!\! \2F1 \l( a,b;\, a+b;\, 1-x \r)\, C_1 
\\
\nonumber
& & 
\!\! + \l( x-1 \r)^{1-b-a}\, \2F1 \l( 1-b,1-a;\, 2-b-a;\, 1-x \r)\, C_2
\end{eqnarray}

When $c$ and $a+b$ are both integers, $g_2$ does not resolve the problem, but
if $a-b$ is not an integer then $g_3$ does, since it permutes the integer
$\lambda = 1-c$ with the non-integer $\kappa = a-b$. For example, for
$a=2/3,\, b=1/3,\, c=1$, \eq{ode_2F1} becomes

\begin{equation}
2\,y/9+ \l( 2\,x-1 \r) \y1+ \l( {x}^{2} -x \r) \yt = 0
\ee

\ni Applying $g_3$ and permuting the parameters $\lambda$ and $\kappa$, we
obtain the following two independent solutions free of integrals

\begin{eqnarray}
y & = &
x^{-1/3}\,
\2F1 \l(1/3,1/3;\,2/3;\,1/x \r)\,
C_1
\\
\nonumber
& & 
+\ 
{x}^{2/3}\,
\2F1\l(2/3,2/3;\,4/3;\,1/x\r)\,
C_2
\end{eqnarray}

When all of $c$, $a+b$ and $a-b$ are integers, these permutations are in
principle of no use, but still for some cases the solution can be represented
free of integrals. This is the case of Legendre's equation. Recalling the
relationship between the associated Legendre function of the first kind and
the hypergeometric \2F1 function\footnote{We use here the Maple convention
for the branch cuts of LegendreP; the idea being discussed is independent of that.},

\begin{eqnarray}
\lefteqn{ {\rm LegendreP} \l( a,b,z \r)\ =} &  &
\\
\nonumber
& & 
{\frac { \l( z+1 \r) ^{1/
2\,b}\2F1 \l( a+1,-a;\, 1-b;\, (1 - z)/2 \r) }{\l( z-1 \r)
^{1/2\,b}\Gamma  \l( 1-b \r) }},
\end{eqnarray}

\ni whenever the group elements of Table 2 can map the \2F1 function
solution into one of the form above, one independent solution can be
expressed using LegendreP and the second one is obtained from the first one
replacing LegendreP by the associated function of the second kind LegendreQ.

%\newpage
% \vfill\eject 

\ni For example, for

\begin{equation}
y/4+ \l( 2\,x -1 \r) \y1+ \l( {x}^{2}-x \r) \yt=0
\ee

\ni we have $\mu = \kappa = \lambda = 0$, so $c=1$ and both $a+b$ and $a-b$
are integers. A solution free of integrals is

\begin{eqnarray}
y & = &{\rm LegendreP} \l( -1/2,2\,x-1 \r)\, C_1
\\
\nonumber
& & 
+\ {\rm LegendreQ} \l( -1/2,2\,x-1 \r)\, C_2 
\end{eqnarray}

%---------------------------------
\section*{Conclusions}

In this presentation we discussed an algorithm for second order linear ODEs,
we called it hyper3, for computing non-Liouvillian solutions by resolving an
equivalence to the \2F1, \1F1 and \0F1 equations. Taking Kamke's book as
testing arena, this algorithm is the most successful one of the current set
of linear ODE algorithms of the Maple system. From the 363 corresponding
examples of Kamke's book having rational coefficients, hyper3 alone solves
331 (91 \%), followed by Kovacic's algorithm solving 181 (50 \%). Moreover,
from these 181 examples admitting Liouvillian solutions, hyper3 solves 163
(90 \%). 

The fact that, for 90\% of these equations admitting Liouvillian solutions,
the solution can also be computed as a hypergeometric one of the form
\eq{sol_pattern} is a good indication that the restriction used to make the
algorithm feasible is appropriate. The fact that around one half of Kamke's
examples only admit special function solutions of non-Liouvillian form also
illustrates the relevance of this type of solution in the general framework
of linear ODE problems popping up in applications.

Despite the simplicity of the approach, till the end of 2001, when the
routines for this algorithm were developed, no equivalent or similar
algorithms were available in any of the Axiom, Maple, Mathematica, MuPAD or
Reduce computer algebra systems (CAS). These CAS failed in computing special
function solutions but for occasional success, e.g., by previous to hyper3 Maple
routines able to resolve an equivalence under {only} power
transformations of the form \eq{power} \cite{george}, or an equivalence
under {only} M\"obius transformations and {only} with respect to the
\2F1 class \cite{mark}. 

Since at the core of hyper3 there is the concept of singularities, two
natural extensions of this work consist of applying the same ideas to
compute solutions for linear ODEs of order three and higher
\cite{MSW_4ODE_challenges} and for second order equations of Heun type. The
latter have four regular singular points or any combination of singularities
derived from that case through confluence processes \cite{slavyanov}; one
example of these are Mathieu equations. Related work is in progress
\cite{mathieu1,mathieu2}.

\medskip
\ni {\bf Acknowledgments}
\medskip

\noindent This work was supported by the MITACS NCE project, the Centre of
Experimental and Constructive Mathematics of Simon Fraser University, and the
NSERC of Canada. We would like to thank one of the referees for her/his kind,
motivating and illustrated comments.

%\vfill\eject 

\end{document}